\documentclass[submission,copyright,creativecommons]{eptcs}
\providecommand{\event}{SOS 2007} 
\usepackage{breakurl}             
\usepackage{underscore}           

\usepackage{algorithm}
\usepackage{algorithmic}

\usepackage{amsmath}
\usepackage{graphicx}
\usepackage{multirow}
\usepackage{wrapfig}

\usepackage{url}%
\usepackage{picinpar}

\usepackage{booktabs}
\usepackage{multirow}

\usepackage{listings}
\lstdefinelanguage{clingo}{
  keywordstyle=[1]\usefont{OT1}{cmtt}{m}{n},%
  keywordstyle=[2]\textbf,%
  keywordstyle=[3]\usefont{OT1}{cmtt}{m}{n},
  alsoletter={\#,\&},%
  keywords=[1]{not,from,import,def,if,else,return,while,break,and,or,for,in,del,and,class},%
  keywords=[2]{\#const,\#show,\#minimize,\#base,\#theory,\#count,\#external,\#program,\#script,\#end,\#heuristic,\#edge,\#project,\#show},%
  keywords=[3]{&,&dom,&sum,&diff,&show,&minimize},%
  morecomment=[l]{\#\ },%
  morecomment=[l]{\%\ },%
  commentstyle={\color{darkgray}}%
}
\newtheorem{definition}{{\bf Definition}}

\newtheorem{example}{\bf Example}

\def\clingo{{\tt clingo}} 
\def\naf{{\: not \: }} 

\title{On Model Reconciliation: \\
How to Reconcile When Robot Does not Know Human's Model?}
\author{
Ho Tuan Dung \qquad\qquad Tran Cao Son
\institute{Department of Computer Science, New Mexico State University, Las Cruces, USA}
\email{\quad dungho@nmsu.edu \quad\qquad stran@nmsu.edu}
}
\def\titlerunning{How to Reconcile When Robot Does not Know Human's Model?}
\def\authorrunning{Ho Tuan Dung \& Tran Cao Son}
\begin{document}
\maketitle

\begin{abstract}
The Model Reconciliation Problem (MRP) was introduced to address issues in explainable AI planning. 
A solution to a MRP is an \emph{explanation} for the differences between the models of the human and the planning agent (robot). Most approaches to solving MRPs assume that the robot, who needs to provide explanations, knows the human model. This assumption is not always realistic in several situations (e.g., the human might decide to update her model and the robot is unaware of the updates).

In this paper, we propose a dialog-based approach for computing explanations of MRPs under the assumptions that ({\em i}) the robot does not know the human model; 
({\em ii}) the human and the robot share the set of predicates of the planning domain and their exchanges are about action descriptions and fluents' values; 
({\em iii}) communication between the parties is perfect; and ({\em iv}) the parties are truthful. A solution of a MRP is computed through a dialog, defined as a sequence of rounds of exchanges, between the robot and the human. In each round, the robot sends a potential  explanation, called \emph{proposal}, to the human who replies with her evaluation of the proposal, called \emph{response}. We develop algorithms for computing proposals by the robot and responses by the human and implement these algorithms in a system that combines imperative means with answer set programming using the multi-shot feature of \clingo{}. 
\end{abstract}

\section{Introduction}

Plan explanation is important for human and AI system (or, robot, as we will refer to the AI system in this paper) to work together in human-aware planning and scheduling. The \emph{model reconciliation problem} (MRP), introduced by \cite{chakraborti2017plan}, is proposed as a way for the robot to explain its solutions to the human. Formally, a MRP is a tuple $(\pi^*, M_R, M_H)$ where $M_R$ is a planning problem of the robot and $M_H$ is an approximation of $M_R$, representing the planning problem that the human considers, and $\pi^*$ is an optimal plan for $M_R$. A solution of a MRP $(\pi^*,M_R,M_H)$, called an \emph{explanation}, is a pair $\epsilon = (\epsilon^+,\epsilon^-)$ such that $M_H^*$, which is obtained from $M_H$ by adding $\epsilon^+$ to it and removing $\epsilon^-$ from it, will have $\pi^*$  as one of its optimal plan.  

Several approaches to solving a MRP have been proposed. 
\cite{chakraborti2017plan} formalized the problem as a search problem and computed an explanation by searching through the space of potential explanations\footnote{A expanded version of this paper is in \cite{SreedharanCK21}. As the key ideas are the same, we will simply refer to \cite{chakraborti2017plan} to save space.}. 
\cite{NguyenSSY20} solved the problem using answer set programming and were able to deal with situations when some actions are missing in the human model or the initial states in the two models are different.   
\cite{Vasileiou21} identified an appropriate explanation by exploiting the existing hitting set duality between minimal correction sets and minimal unsatisfiable sets. 
All of these approaches consider the problem in the context of classical planning and assume that the robot knows the human model $M_H$.
\cite{SreedharanHMK19} argued that it is not always realistic to assume that the robot knows $M_H$ and attempted to address this issue by investigating the \emph{model-free model reconciliation} problem in which the human model is not available in declarative form and not known to the planning agent. Their approach assumes that the robot and human model are represented by Markov Decision Processes (MDP) and the differences between the models are on the transition probabilities, the reward function and the discounting factor. It then applies reinforcement learning to solve the problem.  




In this paper, we present an approach to solving MRPs when the robot \emph{does not know} the human model. 
Different from the model-free model reconciliation problem considered by \cite{SreedharanHMK19}, we assume that the robot and human model are represented by logic programs.  
In our approach, an explanation is computed via a dialog between the robot and the human, which is defined as a sequence of rounds of information exchange. In each round, the robot presents an explanation (about its optimal plan) to the human, who will respond with her evaluation of the proposal. The response of the human will allow the robot to understand the human model. It will also decide whether another round of information exchange is necessary. In this paper, we  assume that the communication between the robot and the human is perfect and the two parties are truthful. Similar to \cite{chakraborti2017plan}, we assume that the human model is an approximation of the robot model. These assumptions ensure that the human and the robot can \emph{understand} each other and the termination of the dialog.  

The main contributions of this paper are a framework for model reconciliation when the robot does not know the human model and its implementation. Specifically, we ({\em i}) define the notions 
of a proposal (from the robot) and a response (from the human); ({\em ii}) develop algorithms for computing proposals and responses; and ({\em iii}) implement the algorithms for computing an explanation via two dialogue controllers, one for the human and one for the robot.










\section{Preliminaries}

\paragraph{Answer set programming (ASP),} introduced by ~\cite{MarekT99} and \cite{Niemela99}, refers to the approach of problem solving using logic programming under the answer set semantics. 
A logic program $\Pi$ is a set of rules of the form \\ 
\hspace*{1cm} $
 a_0  \leftarrow a_1,\ldots,a_m,\naf a_{m+1},\ldots,\naf a_n
$ \\
where $0 \le m \le n$, each $a_i$ is an atom of a propositional language, and 
$\mathit{not}$ represents (default) negation. Intuitively, 
a rule  states that 
if all positive literals               $a_i$ are believed to be true and 
    no negative literal  $\mathit{not}\,a_i$  is believed to be true,
then $a_0$ must be true. 
 Semantically, a logic program induces a set of answer sets, being distinguished models of the program determined by the answer set semantics; see the paper by \cite{GelfondL91} for details.
Answer sets of a program can be computed using the solver \textit{clingo} \cite{GebserKKS14} or \texttt{dlv} \cite{EiterLMPS98}.

\paragraph{Planning Using ASP.}  
 A planning problem---as described using PDDL by \cite{ghallab1998pddl} is a triple $(I,G,D)$, where 
$I$ and $G$ encode the initial state of the world and the goal, respectively; 
and $D$ (the domain) specifies the actions and their preconditions and effects. 
Given a problem $M=(I,G,D)$, we can translate it into a program $\pi(M,n)$  
whose answer sets correspond one-to-one to plans of maximal length $n$ of $M$ \cite{Lifschitz02}.
Standard encoding for computing solutions of planning problems using answer set solvers are available.
In this paper, we will make use of an encoding similar to that 
presented by \cite{NguyenSSY20} that utilizes the translator from 
planning problems in PDDL to logic program facts\footnote{  \url{https://github.com/potassco/plasp}
}  because it simplifies our experiments. 
Program $\pi(M,n)$ consists of different groups of rules:  

\begin{list}{$\bullet$}{\itemsep=0pt \topsep=0pt \parsep=0pt}
\item {\bf Facts:} These atoms define object constants, types of objects, actions, the initial state, and the goal state.

\item {\bf Reasoning About Effects of Actions:} Rules in this group
make sure that an action can only be executed if all of its preconditions are
true and all of the effects of the actions become true. We 
use $h(l,t)$ to denote that  $l$ is true at step $t$ for $1 \le t \le n$.

\item {\bf Goal Enforcement and Action Generation:} To generate action occurrences, we use the choice rule ``$1\{occurs(A,T):action(action(A))\}1 \leftarrow \naf goal(T)$ and to enforce that the goal is satisfied at $n$, we use the rule $\leftarrow \naf goal(n)$ where $goal(i)$ denotes that the goal of the plan satisfies at step $i$.

%

\end{list}  
By computing answer sets of $\pi(M,0),\ldots,\pi(M,k),\ldots$, we can compute optimal plan for $M$. 
It has been proved that if $k$ is the smallest integer such that $\pi(M,k)$ has an answer set $A$ then 
$A$ contains an optimal plan of $M$.


\paragraph{Model Reconciliation Problem.} 
Given two planning problems $M_r = (I_r,G_r,D_r)$ and $M_h = (I_h,G_h,D_h)$, 
a \emph{model reconciliation problem} (MRP) is defined by a tuple $\langle \pi^*, M_r, M_h \rangle$,
where $\pi^*$ is a cost-minimal solution for $M_r$. As in other papers, we define the cost of a plan by its length.
A solution for an MRP is a multi-model explanation $\epsilon=(\epsilon^+,\epsilon^-)$,
which creates a model $M_h^*$ from $M_h$ such that $\pi^*$ is also a cost-minimal 
solution of $M_h^*$ by inserting $\epsilon^+$ to $M_h$ and removing $\epsilon^-$ from $M_h$, where 
$\epsilon^+$ (or $\epsilon^-$) are sets containing some initial conditions, action preconditions/effects, or goals. It is required that the changes in the model of the human must be 
consistent with the robot's model. 

\begin{figwindow}[1,r,%
{\includegraphics[width=0.5\columnwidth]{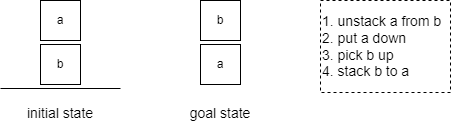}},%
{Achieving $on(b,a)$ given $on(a,b)$ and $on(b,table)$\label{bw-1}}]
\noindent 
{ 
For ease of reading, we will use the well-known Blocksworld domain as running example in this paper. The fluents in this domain are $clear(x)$, $on(x,y)$, $handempty$, $holding(x)$, and $onTable(x)$ and the actions are $unstack(x,y)$, $stack(x,y)$, $pickup(x)$, and $putdown(x)$ with their usual meanings. Let us consider the planning problem depicted in Figure~\ref{bw-1}: the initial state (left) and the goal state consisting of only the literal $on(b,a)$ (right). For space reason, we omit the details of the domain model of the robot and/or the human. It is easy to see that the optimal plan for the robot 
in this setting is $\langle unstack(a,b); putdown(a); pickup(b); stack(b,a) \rangle$.
Assume that $stack(b,a)$ has only one precondition $clear(a)$, i.e., $a$ is clear (which is true in the initial state). In that case, the optimal plan for the human is $\langle  stack(b,a) \rangle$.  
}
\end{figwindow}

\section{Computing Solutions of MRPs via Dialogues}

In this section, we will present our approach for computing a solution of MRPs when the robot does not know the model of planning model of the human ($M_h$) and vice versa. We assume that the two are honest and know that the human model might be incomplete or contain incorrect information. Given that the robot does not know the human model, the only way for the robot to explain to the human the differences between the two problems is to understand the human model through interactions.

\subsection{Formalization: Explanations, Proposals, Responses, and Dialogues}

Informally, the interaction gives the robot and the human the opportunities to inform each other about their models and, ultimately, leads to an agreement. We illustrate this interaction in the following example, where, assuming that the human model does not contain the precondition $holding(x)$ for the action $stack(x,y)$, $clear(x)$ for $pickup(x)$, and  $on(x,y)$ for $unstack(x,y)$, where $x$ and $y$ denote blocks, respectively. 

\begin{example}
[Sample Dialogue]
\label{ex1}
{\rm 
Consider the MRP given by the Blocksworld domain problem in Figure~\ref{bw-1}. 
Intuitively, a dialogue between the robot and the human can be as follows:
\begin{enumerate}
    \item \emph{Robot} (to Human): my optimal plan for the planning problem is $\pi^* = \langle unstack(a,b); putdown(a);$
    $pickup(b);stack(b,a) \rangle$; 
    
    \item \emph{Human} (to Robot): no, my is 
    $\langle stack(b,a) \rangle$ 
    (this is because the \emph{stack}$(b,a)$ action only requires $clear(a)$ as its precondition in the human model); 
    \item \emph{Robot}: 
    your plan is not valid; the \emph{stack}$(b,a)$ action should have the precondition $holding(b)$ (this is because the robot realizes that the condition that prevents the action $stack(b,a)$ to occur in the initial state is $holding(b)$); 
    \item \emph{Human}: 
    ah, okay; in that case, my optimal plan is  
    $\langle pickup(b); stack(b,a) \rangle$ 
    (because the human model does not require $clear(b)$ as a precondition of \emph{pickup}$(b)$); 
 
    \item \emph{Robot}: 
    your plan is still not valid; the \emph{pickup}$(b)$ action should have the precondition $clear(b)$; 
    \item \emph{Human}: 
    I see. In that case, I agree with you. I'll modify the two actions in my model. 
\end{enumerate}
}
\end{example} 
Observe that the robot model still differs from the human model in the action \emph{unstack}. However, the dialogue stops because the optimal plan w.r.t. the human model also has length 4. In this example, the robot has no way to realize that there still exists differences between the two domains even though the human agrees with the robot. This is different in the next example. 
\begin{example}
[Difference in Unstack Must be Removed]
\label{ex2}
\begin{figwindow}[3,r,%
{\includegraphics[width=0.45\columnwidth]{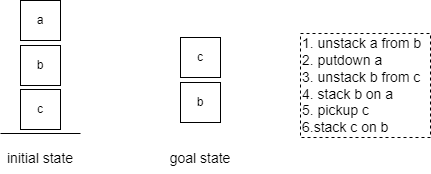}},%
{Getting $c$ on $b$\label{bw-2}}]
\noindent
{
\rm 
Consider the problem depicted in Figure~\ref{bw-2} with similar assumptions as in Example~\ref{ex1}. It is easy to see that  
the sequence $\langle unstack(a,c); putdown(a); unstack(c,b); stack(c,b) \rangle$ is a plan achieving the goal $\{on(c,b)\}$ in the human model, after the human updates the actions \emph{pickup} and \emph{stack} with the correct preconditions as informed by the robot in Steps 3 and 5 in Example~\ref{ex1}. Because this is not a plan with respect to the robot model, the robot will realize that the specification of the action \emph{unstack} in the human model is incorrect, e.g., $unstack(a,c)$ is not executable in the initial state because $on(a,c)$ is a precondition of this action and it is not true in the initial state. It can be easily verified that after the robot informs the human that $unstack(x,y)$ requires $on(x,y)$ and the human updates his/her model with this information, the domain of the human is identical to the domain of the robot. 
}
\end{figwindow}

\end{example}




We will now present our formalization for solving MRP that resembles the dialogue described above. We refer to a message sending from the robot to the human as a \emph{proposal} and a message sending from the human to the  robot  as a \emph{response}. For our definitions, we will need some notations. Let $M = (I,G,D)$ be a planning problem represented as a set of ASP facts as described in the previous section. Let $add(M) = \{add(x) \mid x \in M \}$  
and $remove(M)$ be the set of atoms of the form $delete(x)$ where $x$ is a possible action, 
a precondition,  a postcondition, an initial state atom, or a goal atom. 
We define an explanation as follows. 
\begin{definition}
Given a planning problem $M$, $\epsilon \subseteq add(M) \cup remove(M)$ is called an \emph{explanation} with respect to $M$ if there exists no $x$ such that  $add(x) \in \epsilon$ and $remove(x) \in \epsilon$. 
\end{definition}
Intuitively, an explanation with respect to a planning problem $M$ is given to someone else and consists of elements that should be added/removed with respect to the planning model of the receiver. Given a planning problem $M$ and an explanation $\epsilon$ (w.r.t.  $M'$), we denote with $M \otimes \epsilon = M \setminus \{x \mid remove(x) \in \epsilon\} \cup \{x \mid add(x) \in \epsilon\}$.
%
\begin{definition} 
A \emph{proposal} with respect to a planning problem $M$ is a pair $(\pi, \epsilon)$ where $\pi$ is an optimal solution of $M$ and $\epsilon$ is an explanation with respect to $M$.  
\end{definition} 
Intuitively, each proposal consists of an optimal plan and an explanation\footnote{The first element of the proposal might be unnecessary under the assumption that the planning problem is fixed for the robot. Including the plan will provide the flexibility for the application of the algorithms developed in the next section, e.g., when the initial state or goal are different. We therefore include the plan as a part of an explanation.}. 
In the dialogue between the robot and the human in Example~\ref{ex1}, the first message of the robot to  the human is $(\pi^*,  \emptyset)$; the second one is the pair $(\pi^*, \{add(pre(stack(b,a), holding(b), value(holding(b), true)))\})$; etc. 

Given a proposal from the robot, the human has different options to respond. We need an extra notation for this purpose. 
Let $M$ be a planning problem. Let $\Omega(M)$ be the set of atoms of the form $why(a, t)$ or $why(a, f, v, t)$ where $a$, $f$, $v$, and $t$ is an action, fluent, the truth value of $f$, and a integer ($\le n$), respectively. Intuitively, $why(a, t)$ indicates that action $a$ is not available; and $why(a,f,v,t)$ indicates that action $a$ cannot be executed at step $t$ because $f$ has the value $\neg v$ at this time step. We assume that \emph{goal} is not an action in $M$ and use it as a special action that needs to occur for the goal to be true. 
The notion of a response is defined next. 

\begin{definition} 
\label{def:response}
Given a proposal $(\pi, \epsilon)$, a \emph{response} to $(\pi,\epsilon)$ with respect to a planning problem $M$ is one of the following
\begin{itemize}
    \item $(\top,\top)$, denoting \emph{acceptable} and indicating that 
    $M \otimes \epsilon$ has $\pi$ as one of its optimal plans; 
    \item $(\bot, \epsilon')$, denoting \emph{inapplicable} and indicating that some information in $\epsilon$ is redundant for updating $M$ 
    and $\epsilon' = (add(M) \cap \epsilon) \cup \{remove(x) \mid remove(x) \in \epsilon, x \not\in M\}$; or

    \item $(*, \omega)$, denoting that $\pi$ is \emph{not executable} with respect to $M \otimes \epsilon$ and $\omega \subseteq \Omega(M)$ is a set of atoms explaining why $\pi$ is not executable in   $M \otimes \epsilon$; or 
    \item $(\pi',\top)$ where $\pi'$ is an optimal plan of $M \otimes \epsilon$ 
    and $\pi'$ has smaller cost than $\pi$. 
\end{itemize}

\end{definition} 

Intuitively, a response is constructed given a proposal $\epsilon$ developed with respect to a planning problem $M'$ (i.e., $M'=M_R$) and a planning problem $M$ (i.e., $M = M_H$). It is acceptable if the updated planning problem $M \otimes \epsilon$ has $\pi$ as one of its optimal plans. It is inapplicable if $\epsilon$  is \emph{contradictory} with $M$, i.e., $\epsilon$ contains some $add(x)$ (or $remove(x)$), indicating that $x$ is missing (or redundant) in $M$ but $x$ is indeed in $M$ (or not in $M$). The third option indicates that $\pi$ is invalid in $M \otimes \epsilon$. In this case, $\omega$ provides the robot with the reasons why the human believes that $\pi$ is not executable. The fourth possible response from the human given a proposal $(\pi, \epsilon)$ indicates that the human still has a better plan than $\pi$ after integrating $\epsilon$ into her model. The next example illustrates the third possibility.

\begin{example}
[Plan in Proposal is Not Executable]
{\rm 
For the planning problem given Figure \ref{bw-2}, assume that the human model differs from the robot model in that it does not contain the precondition \emph{handempty} and the postcondition \emph{clear} for the action \emph{unstack}. 

We can easily verify that the plan of the robot $\pi = \langle unstack(a,b), putdown(a), unstack(b,c),$

\noindent $stack(b,a),  pickup(c), stack(c,b)\rangle$ is not executable in the human model because the missing of the postcondition $clear(b)$ of the action $unstack(a,b)$ implies that $clear(b)$ is false after the execution of the sequence $\langle unstack(a,b), putdown(a) \rangle$ (by inertia). Therefore, $unstack(b,c)$ is not executable that implies that $\pi$ is not executable in the human model. In this case, the human will inform the robot that the plan is not executable by responding with $(*, \{ why(unstack(b,c),clear(b),true,3)\})$.




}
\end{example}

\begin{definition}
[Dialogue] 
Given an MRP $(\pi^*, M_R, M_H)$, a dialogue between the robot and the human is a sequence of rounds 
$\langle r_1, \ldots, r_n \rangle$ where, for each $1 \le i \le n$, $r_i = (x_i, y_i)$ and $x_i$ is 
a proposal with respect to $M_r$ and $y_i$ is a response to $x_i$ with respect to $M_H$. 

$\langle r_1, \ldots, r_n \rangle$ is a \emph{successful} dialogue if $y_n = (\top,\top)$. 
\end{definition}



\subsection{Computing Explanations} 

We will now present the algorithms for the robot and human to work together in solving MRPs. 
We assume that the robot and the human are communicating via message passing. 

\subsubsection{Human Dialogue Controller.}

\begin{algorithm}[tb]
\caption{Human Dialogue Controller}
\label{alg:humanController}
\textbf{Input}: Human problem $M_h$ 
    \begin{algorithmic}[1] 
        \STATE $STATUS = \textrm{Running}$
        \WHILE{$STATUS = \textrm{Running}$}
            \STATE Waiting for a proposal $p = (\pi,\epsilon)$
            \IF{$p = \bot$}
                \STATE $STATUS = FAIL$
            \ENDIF
            \STATE $\epsilon'  = \textrm{CheckApplicable}(M_h,\epsilon)$
            \IF {$\epsilon' \neq \emptyset$}
                \STATE $\textrm{Send}(\bot, \epsilon')$
            \ELSE
                \STATE $\widehat{M_h} =  M_h \otimes \epsilon$
                \STATE $\omega = \textrm{ValidatePlan}(\widehat{M_h}, \pi)$
                \IF {$\omega \ne \emptyset$}
                    \STATE $\textrm{Send}(*, \omega)$
                \ELSE
                    \STATE $\pi_h = \textrm{Solve}(\widehat{M_h}, \pi)$
                    \IF {$length(\pi_h) < length(\pi)$} 
                        \STATE $\textrm{Send}(\pi_h, \top)$
                    \ELSE
                        \STATE $\textrm{Send}(\top, \top)$
                        \STATE $STATUS = \textrm{Finish}$
                    \ENDIF
                \ENDIF
            \ENDIF
        \ENDWHILE
    \end{algorithmic}
\end{algorithm}



Algorithm~\ref{alg:humanController} presents the dialogue controller for the human. 
The human is actively waiting for a proposal from the robot (Line 3) and reacts to it following  Definition~\ref{alg:humanController} in computing her responses. The algorithm stops when (\emph{i}) the robot sends $\bot$, indicating that it runs out of suggestion, which means that the problem has no solution, i.e., none of the proposals is acceptable to the human; or (\emph{ii}) the human receives a proposal that is acceptable to her (Line 20). 
The steps in the algorithm are simple and self-explanatory so we omit the details for brevity. We next discuss the procedures that are used in the algorithm. 

\begin{list}{$\bullet$}{\itemsep=0pt \parsep=0pt \topsep=0pt}
    \item {\tt CheckApplicable$(M,\epsilon)$}: this procedure 
    computes  $\epsilon' = (add(M) \cap \epsilon) \cup \{remove(x) \mid remove(x) \in \epsilon, x \not\in M\}$ 
    (this code is simple and we therefore omitted it).


    \item {\tt ValidatePlan($M, \pi$)}: this program checks for the executability condition of the plan $\pi$ given the planning problem $M$. This is done using ASP and the code is included in Listing~\ref{validate}. In this program, $true(x)$ denotes that $x$ is an element in $M$ (e.g., $true(action(a))$ indicates an action in $M$). Given a plan $\pi$ consisting of a set of atoms of the form $occurs(a,i)$, the plan is not executable if ({\em i}) an action in the plan does not exist in $M$ (Lines 14); ({\em ii}) an action is not executable (Lines 15--16); or ({\em iii}) some goal is not satisfied at the end of the plan (Lines 17--18). Note that an action is not executable if one of its preconditions does not hold at the time the action is executed (Lines 10--13).      
\end{list}

\begin{lstlisting}[language=clingo, basicstyle=\small, escapechar=\%,label=validate, caption=ValidatePlan with parameter $M$ and $\pi$]
maxTime(N+1) :- N = #max{T:occurs(_,T)}.
time(1..N-1) :- maxTime(N).
h(X,1) :- initialState(X,value(X,true)).
h(X,T+1) :- time(T), true(action(A)), occurs(A,T), 
 true(postcondition(action(A),
  effect(unconditional),X,value(X,true))).
h(X, T+1) :- time(T),true(action(A)), occurs(A,T), h(X,T), 
  not true(postcondition(action(A),effect(unconditional),X, 
         value(X,false))).
not_executable(A,T,true,X) :- time(T), true(action(A)), 
 true(precondition(action(A),X,value(X,true))), not h(X,T).
not_executable(A,T,false,X) :- time(T), true(action(A)), 
 true(precondition(action(A),X,value(X,false))), h(X,T).
why(A,T) :- time(T), occurs(A,T), not true(action(A)).
why(A,X,V,T) :- time(T), true(action(A)), occurs(A,T), 
 not_executable(A,T,V,X).
why(goal,X,true,N):- maxTime(N),goal(X,value(X,true)),not h(X, N).
why(goal,X,false,N):- maxTime(N),goal(X,value(X,false)),h(X, N). 
\end{lstlisting}


\subsubsection{Robot Dialogue Controller.} The robot dialogue controller is detailed in  Algorithm \ref{alg:robotController}. In order to guide the search for an explanation, 
Algorithm \ref{alg:robotController} uses three variables, $S$, $V$, and $U$, to keep track of information sent to/received from the human. $S$ is a priority queue of proposals that have not been considered where $(\pi, \epsilon)$ has higher priority than $(\pi',\epsilon')$ if $|\epsilon| < |\epsilon'|$. $V$ is a set of proposals that have been considered. $U$ is a set of elements that should not be contained in any proposal which are identified by responses of the form $(\bot, \epsilon')$ from the human. 
Initially, $V=U=\emptyset$ and $S$ is initialized with the trivial proposal $(\pi, \emptyset)$ where $\pi$ is an optimal plan 
(Lines 2--3). Observe that $\pi$ is given in the MRP $(\pi^*,M_r, M_h)$. As such, we might not need to compute the plan $\pi$ in Line 2. 

The robot's main activity is described by a loop (Lines 4--24). In each iteration, it gets a potential proposal from the queue (Line 8), sends to the human (Line 9), waits for a response $r$ (Line 10), and processes the response as follows.   

\begin{itemize}
    \item If $r = (\top,\top)$, by Def.~\ref{def:response} (Case 1), the human accepts the explanation $\epsilon$ and so the dialogue terminates successful (Line 12).
    
    \item If $r = (\bot,\epsilon')$, by Def.~\ref{def:response} (Case 2), $\epsilon'$ should not be contained in the explanation; as such, $U$ is updated (Line 14) and the explanation is dismissed, a potential explanation is then added to $S$ and the dialogue goes into the next round (Lines 15--17).    
 
    \item If $r = (*,\omega)$, by Def.~\ref{def:response} (Case 3), $\omega$ contains information why $\pi$ is not executable in  $M_h \otimes \epsilon$. The robot will therefore expand $\epsilon$ with information from $\omega$ and $U$, update $S$ using Algorithm~\ref{alg:getChangesRPlanFacts} and the dialogue goes into the next round (Line 19).
    
    \item If $r = (\pi',\top)$,  by Def.~\ref{def:response} (Case 4), the robot understands that the human has a plan $\pi'$ that is better than $\pi$. In that case, the robot will identify the reason why $\pi'$ is not an optimal plan in $M_r$ (Line 21), update the queue $S$ (Line 22) using Algorithm~\ref{alg:getChangesHPlanFacts}, and the dialogue moves to the next round.  
\end{itemize}

\begin{algorithm}[tb]
\caption{Robot Dialogue Controller}
\label{alg:robotController}
\textbf{Input}: {Robot problem $M_r$} 
    \begin{algorithmic}[1] 
        \STATE $S = \emptyset, U = \emptyset$, $V = \emptyset $,  $STATUS = \textrm{Running}$
        \STATE $\pi = \textrm{Solve}(M_r)  $
        \STATE $\texttt{enqueue}(S, (\pi, \emptyset))$

        \WHILE{$STATUS = \textrm{Running}$}
            \IF{$S$ is empty}
               \STATE $STATUS = FAIL$ ; Send($\bot$)
            \ENDIF
            \STATE $ (\pi,\epsilon) = \texttt{dequeue}(S)$, $V = V \cup \{(\pi,\epsilon)\}$
            \STATE $\textrm{Send}((\pi,\epsilon))$
            \STATE Waiting for a response $r = (x,y)$ 
            \IF{$x = \top$}
                \STATE $STATUS = \textrm{Finish}$
            \ELSIF{$x = \bot$}
                \STATE $U = U \cup \epsilon'$
                \IF{$(\pi, \epsilon \setminus \epsilon') \not\in V$}
                    \STATE $\texttt{enqueue}(S, (\pi, \epsilon \setminus \epsilon'))$
                \ENDIF
            \ELSIF{$r = (*, \omega)$}
                \STATE $S = updateSR(M_r,\omega,\epsilon,\pi, U, S, V)$
            \ELSIF{$r = (\pi', \top)$}
                \STATE $\omega = \textrm{ValidatePlan}(M_r,\pi')$
                \STATE $S = updateSH(M_r,\omega,\epsilon,\pi, \pi', U, S, V)$
            \ENDIF             
        \ENDWHILE
    \end{algorithmic}
\end{algorithm}

We next discuss the two main procedures in Lines 19 and 22 of Algorithm~\ref{alg:robotController}. 

\medskip
\noindent 
{\bf Algorithm~\ref{alg:getChangesRPlanFacts}:} The algorithm is used in Line 19. 
It computes possible expansions of a proposal $(\pi, \epsilon)$, whose explanation is not rejected by the human, given the responses from the human thus far, i.e., a set of facts indicating the reason why $\pi$ is not executable in the human model ($\omega$) and the set of rejected modifications from the human ($U$).  

In this algorithm, 
$C_1$ is the set of actions that do not occur in the human model and therefore the atom $why(a,t)$ belong to $\omega$ (Line 2--4). 

$C_2$ is the set of missing postconditions that need to be added or redundant preconditions that need to be removed. When a precondition is in the human model and not a part of the robot model and it is required for the action to be executable then it needs to be removed. 
When a precondition is presented in both models and the action is not executable, it means that some action that is supposed to enable this precondition does not contain this postcondition in the human model. Therefore, a postcondition needs to be added. 

\begin{algorithm}[tb]
\caption{updateSR($M_r, \omega, \epsilon, \pi, U, S, V$)}
\label{alg:getChangesRPlanFacts}
\textbf{Input}: $M_r$ - planning problem, $\omega$ - the reason why $\pi$ is not executable, $\pi$ - robot plan, 
$\epsilon$ - robot's proposal, $U$ - redundant information, $S$ - current queue of proposals, $V$ - set of examined proposals\\
\textbf{Output}: Updated queue $S $
    \begin{algorithmic}[1] 
        \STATE $C_1 = C_2 = \emptyset$
        \FOR{each $why(a,t)$ in $\omega$}
            \STATE $C_1 = C_1 \cup  \textrm{AddActions}(M_r, a)$
        \ENDFOR 
        \FOR{each $why(a,x,v,t)$ in $\omega$}
            \IF {$(x,v) \in pre(M_r, a) \vee a=goal$}
                \STATE $C = \textrm{AddPosts}(M_r,\pi,x,v,t)$
                \IF {$C \neq \emptyset$}
                    \STATE $C_2 = C_2 \cup C$
                \ELSE
                    \STATE $C_2 = C_2 \cup \textrm{AddInits}(M_r,x,v) \cup \textrm{RemoveInit}(M_r,x,\neg v)$
                \ENDIF
                 
            \ELSE 
                \STATE $C_2 = C_2 \cup  \textrm{RemovePre}(a,x,v)$
            \ENDIF
        \ENDFOR
        \FOR{each non empty subset $C$ of $C_2 \setminus U$}
            \IF{$(\pi, \epsilon \cup C_1 \cup C) \not\in V$}
                \STATE \texttt{enqueue}$(S, (\pi, \epsilon \cup C_1 \cup C))$
            \ENDIF
        \ENDFOR
        \RETURN $S$
    \end{algorithmic}
\end{algorithm}

The loop (Lines 5--16) is applied to atom of the form $why(a,x, v,t)$ in $\omega$. For each element, 
$C_2$ is updated as follows. For simplicity of the presentation, we write $(x,v)$ is a precondition (resp. a postcondition) of an action $a$ and mean that 
$\textit{pre(action(a), x, value(x,v))}$ 
(resp. $\textit{post(action(a), effect(unconditional),}$
$\textit{x, value(x,v))}$ is an element of the domain model. $pre(M_r,a)$ is the collection of pairs of the form $(x,v)$ that are preconditions of $a$.  
\begin{list}{$\bullet$}{\itemsep=0pt \parsep=0pt \topsep=0pt}
    \item If $why(a,x, v,t) \in \omega$ then $(x,v)$ is a precondition of $a$ in the human model (computed by the code in Listing~\ref{validate}). 
    $(x,v) \in pre(M_r,a)$ (Line 6, $a \ne goal$) indicates that $(x,v)$ is also a precondition of $a$ in the robot model. 
    Therefore, to make $\pi$ executable model in the human model, some action in $\pi$ that occurs before $a$, whose index is less than $t$, must have a postcondition $(x,v)$. 
    This is computed using $AddPosts(M_r, \pi, x, v, t)$. 
    \item If $why(goal,x,v,t) \in \omega$ (Line 6) then $x$ is supposed to have the value $v$ in the goal but it is not true after the execution of $\pi$.   
    Therefore, to achieve the goal and make $\pi$ executable in the human model, 
    some action in $\pi$ must have a postcondition $(x,v)$. 
    This is computed using $AddPosts(M_r, \pi, x, v, t)$. 
    \item If $why(a,x,v,t) \in \omega$ and $(x,v)\not\in pre(M_r,a)$ and $a \ne goal$ then $(x,v)$ is a precondition of $a$ in the human model but it is not a precondition of $a$ in the robot model. Therefore, the precondition should be removed from $a$ (Line 9). 
\end{list}
The updated $S$ (Lines 17--21) includes all possible expansions of $\epsilon$ that have not been considered. Each expansion contains the set of missing actions in the human model ($C_1$) and some potential suggestions from $C_2$.  

The function $AddPosts(M_r,\pi,x,v,t)$ collects actions that occurs before step $t$ that have $(x,v)$ as one of its postconditions and returns this set.
The function $RemovePre(a,x,v)$ returns the precondition that should be removed from $a$. 
When some condition must be true and cannot be established by adding some postcondition or removing some precondition, it can be achieved by adding it to the initial state ($AddInits(M_r,x,v)$, Line 11) 
or removing its contradiction from the initial state ($RemoveInits(M_r,x,\neg v)$, Line 11).  
The codes for these procedures are fairly simple and are omitted for brevity.

\medskip 
\noindent 
{\bf Algorithm~ \ref{alg:getChangesHPlanFacts}:} This algorithm has similar functions as Algorithm~\ref{alg:getChangesRPlanFacts} and is used in Line 22 of Algorithm~\ref{alg:robotController}. The key difference between the two algorithms lies in an additional input, the plan $\pi'$, and therefore, $\omega$. In Algorithm \ref{alg:getChangesHPlanFacts}, $\pi'$ is a plan executable in the human model and has cost smaller than the optimal plan in the robot domain. $\omega$ is the reason why $\pi'$ is not executable in the robot model that is computed using the code in Listing~\ref{validate} with $M_r$ and $\pi'$ as inputs. 
As in the previous algorithm, two sets of changes, $C_1$ and $C_2$, are computed and used to updated the queue of proposals. 

$C_1$ is the set of actions that are not presented in the robot model but occur in the plan from the human model. So, $C_1$ is the set of actions that should be removed (Lines 2--4).  

$C_2$ is the set of changes that prevent $\pi$ to be executable in the human model given the robot's understanding about the human model. The difference between the construction in this algorithm with Algorithm~\ref{alg:getChangesHPlanFacts} is that the robot does not really know what is present in the human model. Therefore, for each $why(a, x, v,t) \in \omega$, the robot must suggest one of the possibilities:
\begin{list}{$\bullet$}{\itemsep=0pt \parsep=0pt \topsep=0pt}
    \item adding the precondition $(x,v)$ for $a$ (Line 8 and Line 10) because this precondition is in the robot model; or
    \item adding the postcondition $(x, \neg v)$ for some action that occurs before $a$ in  $\pi$ (for the case that $(x,v)$ is already a precondition of $a$ in the human model); or 
    \item removing the postcondition $(x, v)$ for some action that occurs before $a$ in  $\pi$; or 
    \item removing some initial state information. 
\end{list}
%

\begin{algorithm}[h]
\caption{updateSH($M_r, \omega, \epsilon, \pi, \pi',  U, S, V$)}
\label{alg:getChangesHPlanFacts}
\textbf{Input}: Similar to Algorithm~\ref{alg:getChangesRPlanFacts} excepts $\pi'$ - a plan from the human model\\
\textbf{Output}: Updated queue $S $
    \begin{algorithmic}[1] 
        \STATE $C_1 = C_2 = \emptyset$
        \FOR{each $why(a,t)$ in $\omega$}
            \STATE $C_1 = C_1 \cup  \textrm{RemoveAction}(a)$
        \ENDFOR 
        \FOR{each $why(a,x,v,t)$ in $\omega$}
            
            \STATE $C = \textrm{AddPosts}(M_r,\pi', x,\neg v,t) \cup \textrm{RemovePost}(M_r,\pi', x,v,t)$
            \IF {$C \neq \emptyset$}
                \STATE $C_2 = C_2 \cup  \textrm{AddPres}(M_r,a,x,v) \cup C $\\
            \ELSE
                \STATE $C_2 = C_2 \cup  \textrm{AddPres}(M_r,a,x,v)   \cup \: \textrm{RemoveInit}(M_r,x,v) $ \\
            \ENDIF
        \ENDFOR
        \FOR{each non empty subset $C$ of $C_2 \setminus U$}
            \IF{$(\pi, \epsilon \cup C_1 \cup C) \not\in V$}
                \STATE \texttt{enqueue}$(S, (\pi, \epsilon \cup C_1 \cup C))$
            \ENDIF
        \ENDFOR
        \RETURN $S$
        
            
                    
    \end{algorithmic}
\end{algorithm}

\subsection{Properties of Algorithms}

It is easy to see that if Algorithms~\ref{alg:humanController}--\ref{alg:robotController} terminate with $STATUS=\textrm{Finish}$ then the two sides agree on a minimal explanation because of the priority in the queue $S$ in Algorithm~\ref{alg:humanController}. It remains to be proven that under the assumption that $M_R$ has a solution then the algorithms will terminate with $STATUS=\textrm{Finish}$. This can be achieved using the following observations: 
\begin{list}{$\bullet$}{\itemsep=0pt \parsep=0pt \topsep=0pt}
  \item Algorithm~\ref{alg:robotController} terminates with $STATUS=\textrm{FAIL}$ only if the queue $S = \emptyset$. 
  \item Given that $(\pi,\epsilon)$ is a proposal sent by the robot to the human and the response from the human is either $(*,\omega')$ or $ (\pi',\top)$ then  
  Algorithm~\ref{alg:getChangesRPlanFacts} or \ref{alg:getChangesHPlanFacts} adds at least one proposal $(\pi, \tau)$ to $S$
  such that $\epsilon \subset \tau$ and for every $add(x) \in \tau$ (resp. $remove(x)$), $x \in M_H$ (resp. $x \not\in M_H$); 
  \item the response from the human for the first element of $S$, $(\pi, \epsilon)$, is of the form  $(*,\omega')$ or $ (\pi',\top)$. 
\end{list}

\section{Experimental Evaluation}

We implement the algorithms described in the previous section and empirically validate the system using the benchmarks that have been used in comparing different systems that solve MRPs. Observe that because we solve MRPs under the assumption that the robot does not know the human model, a comparison on the efficiency between our system and others would not be a fair one. Nevertheless, a comparison between the output would be a validation of our approach. 

We run our system using the same benchmarks used by \cite{chakraborti2017plan}: the three planning domains \textit{LOGISTICS, BLOCKSWORLD} and \textit{ROVER} with the following setting. We use the problem instances as given, the robot model $D_r$, and create $D_h$ from $D_r$ by \textbf{(1)} removing one random precondition for every action in $D_r$; \textbf{(2)} removing one random postcondition for every action in $D_r$; \textbf{(3)} removing one random precondition and one random postcondition for every action in $D_r$;  \textbf{(4)} removing all but one random precondition and one random postcondition from two actions in $D_r$; \textbf{(5)} removing some random actions that used in the robot plan; and \textbf{(6)} modifying initial states. Exceptions are made when an action has only one precondition/postcondition.


\begin{table}[h!]
 \centering
 {\tiny
    \begin{tabular}{@{\extracolsep{\fill}}|cr|c||cccccc|cccccc|cccccc|}
    \toprule
    \multicolumn{2}{|c|}{\multirow{2}{*}{Problem}} &
    \multicolumn{1}{|c||}{\multirow{2}{*}{$|\pi|$}} & 
    \multicolumn{6}{c|}{Mod. 1} & 
    \multicolumn{6}{c|}{Mod. 2} & 
    \multicolumn{6}{c|}{Mod. 3}  \\
    
    \multicolumn{2}{|c|}{} &
    \multicolumn{1}{|c||}{}& 
    $|\epsilon|$ & {$r$} & {$T$} & $t$ & $T_C$ & $T_N$ & 
    $|\epsilon|$ & {$r$} & {$T$} & $t$ & $T_C$ & $T_N$ & 
    $|\epsilon|$ & {$r$} & {$T$} & $t$ & $T_C$ & $T_N$ \\
    \hline
    \hline
    \multirow{4}{*}{\rotatebox{90}{\hspace{0.0em}\textsc{Blocks-}} \rotatebox{90}{\hspace{0.0em}\textsc{world}}} 
& $i$   & 16 
        & 3  & 4 & 155 & 2  & 6  & 88   
        & 4  & 6 & 111 & 3  & 0.3  & 34 
        & 7  & 6 & 245 & 6  & 44  & 37 
        \\
        & $ii$  & 20 
        & 3  & 3 & 58 & 1  & 43  & 81  
        & 3  & 4 & 27 & 3  & 0.5  & 21  
        & 7  & 6 & 33 & 5  & 31  & 27
        \\
& $iii$ & 18 
        & 3  & 4 & 348 & 2  & 3  & 290  
        & 4  & 5 & 394 & 4  & 0.5  & 222  
        & 7  & 5 & 239 & 5  & 30  & 223  
        \\
& $iv$  & 20 
        & 3  & 3 & 360  & 2 & 13  & 221  
        & 3  & 2 & 301  & 3 & 0.7  & 280 
        & 8  & 6 & 4124 & 4 & 23  & 310
        \\
\hline
\multirow{4}{*}{\rotatebox{90}{\hspace{0.0em}\textsc{Logis-}} \rotatebox{90}{\hspace{0.0em}\textsc{tics}}} 
& $v$ & 20 
    & 5  & 5 & 145 & 3 & 19  & 117
    & 3  & 2 & 256 & 5 & 0.6  & 113 
    & 10 & 6 & 149 & 8 & 459 & 145 
    \\
& $vi$ & 17 
    & 5  & 4 & 161 & 3  & 21  & 103
    & 4  & 4 & 133 & 4  & 5  & 99  
    & 10 & 5 & 216 & 7  & 399 & 110 
    \\
& $vii$ & 14 
    & 5  & 4 & 20 & 2  & 20 & 1397
    & 4  & 4 & 22 & 4  & 3  & 1167 
    & 10 & 6 & 28 & 6  & 397 & 1173
    \\
& $viii$ & 24 
    & 5  & 4 & 828  & 2  & 23 & 18    
    & 4  & 4 & 1287 & 6  & 0.7 & 17    
    & 10 & 6 & 805  & 10 & 468 & 20
    \\
\hline
\multirow{4}{*}{\rotatebox{90}{\hspace{0.0em}\textsc{Rover}}} 
& $ix$ & 10 
    & 5 & 4 & 7  & 2   & 46 & 7
    & 5 & 4 & 14 & 8   & 6 & 5  
    & 5 & 3 & 10 & 4   & 570 & 6  
    \\
& $x$ & 8 
    & 2 & 2 & 2  & 1   & 2 & 3
    & 6 & 2 & 6  & 3   & 7 & 3  
    & 4 & 5 & 5  & 3   & 374 & 4  
    \\
& $xi$ & 11
    & 6& 3 & 15 & 5   & 49 & 11  
    & 6& 4 & 26 & 16  & 11 & 15  
    & 8& 6 & 30 & 13  & 569 & 40 
    \\
& $xii$ & 8 
    & 4& 3 & 10 & 5  & 3 & 2
    & 4& 4 & 16 & 8  & 4 & 11  
    & 5& 4 & 24 & 12 & 61 & 17  \\
\hline 
\hline 
    \end{tabular}
    }
\label{tab:exp1}
\end{table}

\begin{table}[h!]
 \centering
 {\tiny
    \begin{tabular}{@{\extracolsep{\fill}}|cr|c||cccccc|ccccc|ccccc|}
    \toprule
    \multicolumn{2}{|c|}{\multirow{2}{*}{Problem}} &
    \multicolumn{1}{|c||}{\multirow{2}{*}{$|\pi|$}} & 
    \multicolumn{6}{c|}{Mod. 4} & 
    \multicolumn{5}{c|}{Mod. 5} & 
    \multicolumn{5}{c|}{Mod. 6} \\
    
    \multicolumn{2}{|c|}{} &
    \multicolumn{1}{|c||}{}& 
    $|\epsilon|$ & {$r$} & {$T$} & $t$ & $T_C$ & $T_N$ &
    $|\epsilon|$ & {$r$} & {$T$} & $t$ & $T_N$ &
    $|\epsilon|$ & {$r$} & {$T$} & $t$ & $T_N$ \\
    \hline
    \hline
    \multirow{4}{*}{\rotatebox{90}{\hspace{0.0em}\textsc{Blocks-}} \rotatebox{90}{\hspace{0.0em}\textsc{world}}} 
& $i$   & 16 
        & 11 & 7 & 606 & 12 & 452 & 81 
        & 9  & 1 & 69  & 5  & 47
        & 2  & 1 & 3  & 1  & 1
        \\
        & $ii$  & 20 
        & 9  & 6 & 27 & 6  & 437  & 93 
        & 20 & 3 & 29 & 10 & 43
        & 4 & 3 & 5 & 3 & 1
        \\
& $iii$ & 18 
        & 8  & 5 & 107 & 9  & 652  & 277  
        & 18 & 3 & 218 & 12 & 45
        & 5 & 3 & 7 & 2 & 3
        \\
& $iv$  & 20 
        & 8  & 6 & 344  & 8 & 473  & 213 
        & 26 & 2 & 418  & 8 & 46
        & 5 & 2 & 41  & 19 & 99
        \\
\hline
\multirow{4}{*}{\rotatebox{90}{\hspace{0.0em}\textsc{Logis-}} \rotatebox{90}{\hspace{0.0em}\textsc{tics}}} 
& $v$ & 20 
    & 4  & 3 & 160 & 2 & 16 & 137  
    & 4  & 3 & 135 & 8 & 56
    & 5  & 2 & 237 & 57 & 112
    \\
& $vi$ & 17 
    & 4  & 4 & 234 & 3  & 15  & 103 
    & 9  & 3 & 117 & 10 & 57
    & 4  & 3 & 221 & 97 & 124
    \\
& $vii$ & 14 
    & 4  & 3 & 19 & 2  & 14  & 1213  
    & 19 & 3 & 24 & 9  & 59
    & 7 & 3 & 278 & 49  & 121
    \\
& $viii$ & 24 
    & 4  & 3 & 953  & 2  & 14 & 21    
    & 16 & 3 & 670  & 13 & 59
    & 6 & 2 & 283  & 102 & 132
    \\
\hline
\multirow{4}{*}{\rotatebox{90}{\hspace{0.0em}\textsc{Rover}}} 
& $ix$ & 10 
    & 8 & 4 & 18 & 6   & 589 & 7  
    & 41 & 3 & 25 & 23 & 323
    & 4 & 2 & 10 & 23 & 3
    \\
& $x$ & 8 
    & 9 & 5 & 15 & 6   & 1513 & 3
    & 35 & 3 & 14 & 13 & 354
    & 4 & 2 & 8 & 13 & 1
    \\
& $xi$ & 11
    & 11 & 6 & 48 & 18& 1516 & 14 
    & 21 & 3 & 61 & 53& 1498
    & 6 & 3 & 71 & 53& 15
    \\
& $xii$ & 8 
    & 8& 3 & 42 & 12 & 39 & 9
    & 29 & 3 & 66 & 62& 2537
    & 10 & 4 & 56 & 62& 18 \\
\hline 
\hline 
    \end{tabular}
    }
\caption{Varying Modifications and Domains}
\end{table}

Table \ref{tab:exp1} tabulates the optimal plan lengths $\pi$, explanation lengths $|\epsilon|$ and run-time in seconds. In our proposed framework, $r$, $T$, $t$ is number of rounds, the total running time, and the running time excluding the planning time in both agents, respectively. For the two other framework, we record $T_C$ - the total runtime of the framework $C$ (the system by \cite{chakraborti2017plan}) and $T_N$ (the system by \cite{NguyenSSY20}).  Observe that we do not experiment with the system by \cite{Vasileiou21} as this system is similar to the system $C$ (or $N$), just faster. 

Examining the outputs confirms that our system is correct as it returns the same answers returned by the system $N$ \cite{NguyenSSY20} or the system $C$ \cite{chakraborti2017plan}. It is interesting to observe that the number of rounds required for the robot to find a solution ($r$) is not proportional to the size of the explanation ($|\epsilon|$) (see, e.g., Mod. 1. ROVER, \emph{iv} vs. \emph{xi}). 
As expected, our system is slower than other systems. As it can be seen, the major time spending in our system is for planning since several calls to the planner (Solve) need to be made during the course of a dialogue. 



\section{Conclusions, Discussion, and Future Work}

We propose a dialogue-based framework for solving MRPs that differs from earlier approaches to solving MRPs  \cite{chakraborti2017plan,NguyenSSY20,Vasileiou21} in that our framework does not assume that the robot, who needs to solve the MRPs, knows the human model. 
We develop the notions of a proposal/response from the robot/human and algorithms for computing proposals/responses as well as for controlling of the dialogue between the robot and the human. 
We implement the proposed algorithms in a distributed setting and validate the usability of our system using benchmarks from the literature. Our experiments show that the proposed system computes correct answers but is not as efficient as state-of-the-art MRP solvers. Improving the performance of the system by identifying better strategy for  selecting proposal (e.g., aiming at a faster convergence of the dialogue) than the current brute-force strategy could be a worthy undertaking that we leave for the future. 

To the best of our knowledge, the work that is most closely to our framework is proposed by \cite{abs-2011-12262}. In this work, a method for generating questions to refine the robot's understanding of the human model is proposed. The interaction between the agents is also dialog-based but the content of the dialog is different. In each interaction, the robot provides a tuple $(I,G,\pi)$, the initial state, goal state and a plan, and the answer from the human is simply $yes$/$no$. As such, \cite{abs-2011-12262} focuses on identifying sequence of questions that help the robot to learn the human model. The work of \cite{SreedharanHMK19} also removes the assumption that the robot knows the human model but employs reinforcement learning to solve the problem.  

Finally, we note that the use of ASP as the representation language for planning in solving MRPs implies that the proposed framework can be used for computing solutions of the Model Reconciliation in Logic Programs (MRLP) problems proposed by \cite{SonNV021}.

\section*{Acknowledgement}

The authors have been partially supported by NSF grants 1914635, 1757207, and 1812628.

\bibliographystyle{eptcs}
\bibliography{ref,bib2010,bibfile}

\newpage

\section{Appendix 1: Algorithms}

This appendix contains the descriptions of some algorithms that are referred to in the paper but are omitted due to lack of space. Algorithm~\ref{alg:addposts} identifies potential postconditions of actions in the plan $\pi$ that occur before step $t$  and assigns the variable $x$ the truth value $v$. Note that in our representation, actions and fluents are predicates. As such, if a postcondition is added to an action $a$ then it is reasonable to add a similar postcondition to an action that is unifiable with $a$. Line 5 of Algorithm~\ref{alg:addposts} ensures that this is done.



\begin{algorithm}[h]
\caption{addPosts($M_r, \pi, x, v, t$)}
\label{alg:addposts}
\textbf{Input}: $M_r$ - planning problem, $\pi$ - a plan, $a$ - action, $x$ - fluent, $v$ - truth value of the fluent, $t$ - a time step\\
\textbf{Output}: $\epsilon$ - set of elements to be added 
    \begin{algorithmic}[1] 
        \STATE $\epsilon = \emptyset$
        \FOR{$i = t-1$ to $1$}
            \STATE $a = \pi[i]$
            \FOR{each $p = postcondition(a',\textrm{effect},x',v') \in M_r$}
                \IF{$sameName(a,a')$ and $sameName(x,x')$ and $v = v'$}
                    \STATE $\epsilon = \epsilon \cup \{add(p)\}$
                \ENDIF
            \ENDFOR
        \ENDFOR
        \RETURN $\epsilon$
    \end{algorithmic}
\end{algorithm}

Algorithm~\ref{alg:addInit} checks whether the variable $x$ has the value $v$ in the initial state of $M_r$ and returns the set of elements that need to be added to the initial state (of the human). 

\begin{algorithm}[h]
\caption{addInit($M_r, x, v$)}
\label{alg:addInit}
\textbf{Input}: $M_r$ - planning problem, $x$ - fluent, $v$ - truth value of the fluent\\
\textbf{Output}: $\epsilon$ - initial state modification 
    \begin{algorithmic}[1] 
        \STATE $\epsilon = \emptyset$
        \IF{$init(x,v) \in M_r$}
            \STATE $\epsilon = \epsilon \cup \{add(init(x,v))\}$
        \ENDIF
        \RETURN $\epsilon$
    \end{algorithmic}
\end{algorithm}


Algorithm~\ref{alg:addPres} checks whether the action $a$ has $(x,v)$ as one of its precondition in $M_r$ and returns the set of preconditions need to be added to the human model. This algorithm also consider  actions that are unifiable with $a$ as in Algorithm~\ref{alg:addposts}.

\begin{algorithm}[h]
\caption{addPres($M_r, a, x, v$)}
\label{alg:addPres}
\textbf{Input}: $M_r$ - planning problem, $a$ - an action, $x$ - a fluent, $v$ - truth value of the fluent\\
\textbf{Output}: $\epsilon$ - set of preconditions that should be added 
    \begin{algorithmic}[1] 
        \STATE $\epsilon = \emptyset$
        \FOR{each $p = precondition(a',x',v') \in M_r$}
            \IF{$sameName(a,a')$ and $sameName(x,x')$ and $v = v'$}
                \STATE $\epsilon = \epsilon \cup \{ add(p) \}$
            \ENDIF
        \ENDFOR
        \RETURN $\epsilon$
    \end{algorithmic}
\end{algorithm}

Algorithm~\ref{alg:removePost} identifies postconditions of the form $(x,v)$ of actions in the plan $\pi$ that occur before step $t$ that are not in the robot model, and thus, have to be removed.  
  
\begin{algorithm}[h]
\caption{removePost($M_r, \pi, x, v, t$)}
\label{alg:removePost}
\textbf{Input}: $M_r$ - planning problem, $\pi$ - a plan, $x$ - a fluent, $v$ - truth value of the fluent, $t$ - a time step\\
\textbf{Output}: $\epsilon$ - set of postconditions to be removed 
    \begin{algorithmic}[1] 
        \STATE $\epsilon = \emptyset$
        \FOR{$i = t-1$ to $1$}
            \STATE $a = \pi[i]$
            \STATE $p = postcondition(a,\textrm{effect},x,v)$
            \IF{$p \not\in M_r$}
                \STATE $\epsilon = \epsilon \cup \{ remove(p)\}$
            \ENDIF
        \ENDFOR
        \RETURN $\epsilon$
    \end{algorithmic}
\end{algorithm}

Algorithm~\ref{alg:removeInit} checks whether the variable $x$ has the value $v$ in the initial state of $M_r$ and returns the set of elements that need to be removed from the initial state (of the human model). 

\begin{algorithm}[h]
\caption{removeInit($M_r, x, v,$)}
\label{alg:removeInit}
\textbf{Input}: $M_r$ - planning problem, $x$ - a fluent, $v$ - truth value of the fluent\\
\textbf{Output}: The explanation $\epsilon$
    \begin{algorithmic}[1] 
        \STATE $\epsilon = \emptyset$
        \IF{$i = init(x,v) \not\in M_r$}
            \STATE $\epsilon = \epsilon \cup \{ remove(i)\}$
        \ENDIF
        \RETURN $\epsilon$
    \end{algorithmic}
\end{algorithm}


Algorithm~\ref{alg:removePre} returns the set of precondition of $a$ that needs to be removed from the human model. 

\begin{algorithm}[h]
\caption{removePre($a,x,v$)}
\label{alg:removePre}
\textbf{Input}: $a$ - an action, $x$ - a fluent, $v$ - truth value of the fluent\\
\textbf{Output}: $\epsilon$ -- set of preconditions to be removed 
    \begin{algorithmic}[1] 
        \RETURN $\{ remove(precondition(a,x,v))\}$
    \end{algorithmic}
\end{algorithm}

Algorithm~\ref{alg:sameName} returns true if two terms are unifiable and false otherwise.

\begin{algorithm}[h]
\caption{sameName($t_1,t_2$)}
\label{alg:sameName}
\textbf{Input}: $t_1,t_2$ are terms presenting actions or fluents\\
\textbf{Output}: true/false
    \begin{algorithmic}[1] 
        \STATE $s_1 = getSignature(t_1)$
        \STATE $s_2 = getSignature(t_2)$
        \RETURN $s_1 = s_2$
    \end{algorithmic}
\end{algorithm}

\newpage

\section{Appendix 2: Blocksworld Domain - PDDL and ASP Encoding}

Listing~\ref{listdomain} is the PDDL encoding of the Blocksworld domain used in the paper and the experiment. 


\begin{lstlisting}[language=clingo, basicstyle=\small, escapechar=\%,caption=Robot Domain in PDDL,label=listdomain] 
(define (domain BLOCKS)
  (:requirements :strips :typing)
  (:types block)
  (:predicates (on ?x-block ?y-block)
   (ontable ?x-block) (clear ?x-block)
   (handempty) (holding ?x-block)
  )
  (:action pickup
   :parameters (?x-block)
   :precondition(and  %\underline{(clear ?x)}% 
     (ontable ?x) (handempty))
   :effect (and (not (ontable ?x))
	 (not (clear ?x))
	 (not (handempty)) (holding ?x)))
  (:action putdown
   :parameters (?x-block)
   :precondition (holding ?x) 
   :effect  (and (not (holding ?x))
    (clear ?x) (handempty) (ontable ?x)))
  (:action stack
   :parameters (?x-block ?y-block)
   :precondition (and %\underline{(holding ?x)}% (clear ?y)) 
   :effect (and (not (holding ?x))
	(not (clear ?y)) (clear ?x)
	(handempty) (on ?x ?y)))
  (:action unstack
   :parameters (?x-block ?y-block)
   :precondition (and %\underline{(on ?x ?y)}% 
    (clear ?x) (handempty))
   :effect (and (holding ?x) (clear ?y)
    (not (clear ?x)) (not (handempty))
	(not (on ?x ?y)))))
\end{lstlisting} 

The next listing represents the set of facts encoding a planning problem in the Blocksworld domain for the robot 
in Figure~\ref{bw-1}.

\begin{lstlisting}[language=clingo, basicstyle=\small, escapechar=\%,caption=The Robot Planning Problem as ASP facts,label=aspfacts] 
constant(c("b")).
constant(c("a")).

has(c("b"),type("block")).
has(c("a"),type("block")).

initialState(var("handempty"),value(var("handempty"),true)).
initialState(var(("clear",c("a"))),
    value(var(("clear",c("a"))),true)).
initialState(var(("on",c("a"),c("b"))),
    value(var(("on",c("a"),c("b"))),true)).
initialState(var(("ontable",c("b"))),
    value(var(("ontable",c("b"))),true)).

action(action(("unstack",c("b"),c("b")))).
pre(action(("unstack",c("b"),c("b"))),var("handempty"),
    value(var("handempty"),true)).
pre(action(("unstack",c("b"),c("b"))),var(("clear",c("b"))),
    value(var(("clear",c("b"))),true)).
pre(action(("unstack",c("b"),c("b"))),var(("on",c("b"),c("b"))),
    value(var(("on",c("b"),c("b"))),true)).
post(action(("unstack",c("b"),c("b"))),
    effect(unconditional),var(("on",c("b"),c("b"))),
    value(var(("on",c("b"),c("b"))),false)).
post(action(("unstack",c("b"),c("b"))),
    effect(unconditional),var(("clear",c("b"))),
    value(var(("clear",c("b"))),false)).
post(action(("unstack",c("b"),c("b"))),
    effect(unconditional),var("handempty"),
    value(var("handempty"),false)).
post(action(("unstack",c("b"),c("b"))),
    effect(unconditional),var(("clear",c("b"))),
    value(var(("clear",c("b"))),true)).
post(action(("unstack",c("b"),c("b"))),
    effect(unconditional),var(("holding",c("b"))),
    value(var(("holding",c("b"))),true)).


action(action(("unstack",c("a"),c("b")))).
pre(action(("unstack",c("a"),c("b"))),var("handempty"),
    value(var("handempty"),true)).
pre(action(("unstack",c("a"),c("b"))),var(("clear",c("a"))),
    value(var(("clear",c("a"))),true)).
pre(action(("unstack",c("a"),c("b"))),var(("on",c("a"),c("b"))),
    value(var(("on",c("a"),c("b"))),true)).
post(action(("unstack",c("a"),c("b"))),
    effect(unconditional),var(("on",c("a"),c("b"))),
    value(var(("on",c("a"),c("b"))),false)).
post(action(("unstack",c("a"),c("b"))),
    effect(unconditional),var("handempty"),
    value(var("handempty"),false)).
post(action(("unstack",c("a"),c("b"))),
    effect(unconditional),var(("clear",c("a"))),
    value(var(("clear",c("a"))),false)).
post(action(("unstack",c("a"),c("b"))),
    effect(unconditional),var(("clear",c("b"))),
    value(var(("clear",c("b"))),true)).
post(action(("unstack",c("a"),c("b"))),
    effect(unconditional),var(("holding",c("a"))),
    value(var(("holding",c("a"))),true)).

action(action(("unstack",c("b"),c("a")))).
pre(action(("unstack",c("b"),c("a"))),var("handempty"),
    value(var("handempty"),true)).
pre(action(("unstack",c("b"),c("a"))),var(("clear",c("b"))),
    value(var(("clear",c("b"))),true)).
pre(action(("unstack",c("b"),c("a"))),var(("on",c("b"),c("a"))),
    value(var(("on",c("b"),c("a"))),true)).
post(action(("unstack",c("b"),c("a"))),
    effect(unconditional),var(("on",c("b"),c("a"))),
    value(var(("on",c("b"),c("a"))),false)).
post(action(("unstack",c("b"),c("a"))),
    effect(unconditional),var("handempty"),
    value(var("handempty"),false)).
post(action(("unstack",c("b"),c("a"))),
    effect(unconditional),var(("clear",c("a"))),
    value(var(("clear",c("a"))),true)).
post(action(("unstack",c("b"),c("a"))),
    effect(unconditional),var(("clear",c("b"))),
    value(var(("clear",c("b"))),false)).
post(action(("unstack",c("b"),c("a"))),
    effect(unconditional),var(("holding",c("b"))),
    value(var(("holding",c("b"))),true)).

action(action(("unstack",c("a"),c("a")))).
pre(action(("unstack",c("a"),c("a"))),var("handempty"),
    value(var("handempty"),true)).
pre(action(("unstack",c("a"),c("a"))),var(("clear",c("a"))),
    value(var(("clear",c("a"))),true)).
pre(action(("unstack",c("a"),c("a"))),var(("on",c("a"),c("a"))),
    value(var(("on",c("a"),c("a"))),true)).
post(action(("unstack",c("a"),c("a"))),
    effect(unconditional),var(("on",c("a"),c("a"))),
    value(var(("on",c("a"),c("a"))),false)).
post(action(("unstack",c("a"),c("a"))),
    effect(unconditional),var("handempty"),
    value(var("handempty"),false)).
post(action(("unstack",c("a"),c("a"))),
    effect(unconditional),var(("clear",c("a"))),
    value(var(("clear",c("a"))),false)).
post(action(("unstack",c("a"),c("a"))),
    effect(unconditional),var(("clear",c("a"))),
    value(var(("clear",c("a"))),true)).
post(action(("unstack",c("a"),c("a"))),
    effect(unconditional),var(("holding",c("a"))),
    value(var(("holding",c("a"))),true)).

action(action(("stack",c("b"),c("b")))).
pre(action(("stack",c("b"),c("b"))),var(("clear",c("b"))),
    value(var(("clear",c("b"))),true)).
pre(action(("stack",c("b"),c("b"))),var(("holding",c("b"))),
    value(var(("holding",c("b"))),true)).
post(action(("stack",c("b"),c("b"))),
    effect(unconditional),var(("on",c("b"),c("b"))),
    value(var(("on",c("b"),c("b"))),true)).
post(action(("stack",c("b"),c("b"))),
    effect(unconditional),var("handempty"),
    value(var("handempty"),true)).
post(action(("stack",c("b"),c("b"))),
    effect(unconditional),var(("clear",c("b"))),
    value(var(("clear",c("b"))),true)).
post(action(("stack",c("b"),c("b"))),
    effect(unconditional),var(("clear",c("b"))),
    value(var(("clear",c("b"))),false)).
post(action(("stack",c("b"),c("b"))),
    effect(unconditional),var(("holding",c("b"))),
    value(var(("holding",c("b"))),false)).

action(action(("stack",c("a"),c("b")))).
pre(action(("stack",c("a"),c("b"))),var(("clear",c("b"))),
    value(var(("clear",c("b"))),true)).
pre(action(("stack",c("a"),c("b"))),var(("holding",c("a"))),
    value(var(("holding",c("a"))),true)).
post(action(("stack",c("a"),c("b"))),
    effect(unconditional),var(("on",c("a"),c("b"))),
    value(var(("on",c("a"),c("b"))),true)).
post(action(("stack",c("a"),c("b"))),
    effect(unconditional),var("handempty"),
    value(var("handempty"),true)).
post(action(("stack",c("a"),c("b"))),
    effect(unconditional),var(("clear",c("a"))),
    value(var(("clear",c("a"))),true)).
post(action(("stack",c("a"),c("b"))),
    effect(unconditional),var(("clear",c("b"))),
    value(var(("clear",c("b"))),false)).
post(action(("stack",c("a"),c("b"))),
    effect(unconditional),var(("holding",c("a"))),
    value(var(("holding",c("a"))),false)).


action(action(("stack",c("b"),c("a")))).
pre(action(("stack",c("b"),c("a"))),var(("clear",c("a"))),
    value(var(("clear",c("a"))),true)).
pre(action(("stack",c("b"),c("a"))),var(("holding",c("b"))),
    value(var(("holding",c("b"))),true)).
post(action(("stack",c("b"),c("a"))),
    effect(unconditional),var(("on",c("b"),c("a"))),
    value(var(("on",c("b"),c("a"))),true)).
post(action(("stack",c("b"),c("a"))),
    effect(unconditional),var("handempty"),
    value(var("handempty"),true)).
post(action(("stack",c("b"),c("a"))),
    effect(unconditional),var(("clear",c("a"))),
    value(var(("clear",c("a"))),false)).
post(action(("stack",c("b"),c("a"))),
    effect(unconditional),var(("clear",c("b"))),
    value(var(("clear",c("b"))),true)).
post(action(("stack",c("b"),c("a"))),
    effect(unconditional),var(("holding",c("b"))),
    value(var(("holding",c("b"))),false)).


action(action(("stack",c("a"),c("a")))).
pre(action(("stack",c("a"),c("a"))),var(("clear",c("a"))),
    value(var(("clear",c("a"))),true)).
pre(action(("stack",c("a"),c("a"))),var(("holding",c("a"))),
    value(var(("holding",c("a"))),true)).
post(action(("stack",c("a"),c("a"))),
    effect(unconditional),var(("on",c("a"),c("a"))),
    value(var(("on",c("a"),c("a"))),true)).
post(action(("stack",c("a"),c("a"))),
    effect(unconditional),var("handempty"),
    value(var("handempty"),true)).
post(action(("stack",c("a"),c("a"))),
    effect(unconditional),var(("clear",c("a"))),
    value(var(("clear",c("a"))),true)).
post(action(("stack",c("a"),c("a"))),
    effect(unconditional),var(("clear",c("a"))),
    value(var(("clear",c("a"))),false)).
post(action(("stack",c("a"),c("a"))),
    effect(unconditional),var(("holding",c("a"))),
    value(var(("holding",c("a"))),false)).

action(action(("putdown",c("b")))).
pre(action(("putdown",c("b"))),var(("holding",c("b"))),
    value(var(("holding",c("b"))),true)).
post(action(("putdown",c("b"))),
    effect(unconditional),var(("ontable",c("b"))),
    value(var(("ontable",c("b"))),true)).
post(action(("putdown",c("b"))),
    effect(unconditional),var("handempty"),
    value(var("handempty"),true)).
post(action(("putdown",c("b"))),
    effect(unconditional),var(("clear",c("b"))),
    value(var(("clear",c("b"))),true)).
post(action(("putdown",c("b"))),
    effect(unconditional),var(("holding",c("b"))),
    value(var(("holding",c("b"))),false)).

action(action(("putdown",c("a")))).
pre(action(("putdown",c("a"))),var(("holding",c("a"))),
    value(var(("holding",c("a"))),true)).
post(action(("putdown",c("a"))),
    effect(unconditional),var(("ontable",c("a"))),
    value(var(("ontable",c("a"))),true)).
post(action(("putdown",c("a"))),
    effect(unconditional),var("handempty"),
    value(var("handempty"),true)).
post(action(("putdown",c("a"))),
    effect(unconditional),var(("clear",c("a"))),
    value(var(("clear",c("a"))),true)).
post(action(("putdown",c("a"))),
    effect(unconditional),var(("holding",c("a"))),
    value(var(("holding",c("a"))),false)).

action(action(("pickup",c("b")))).
pre(action(("pickup",c("b"))),var("handempty"),
    value(var("handempty"),true)).
pre(action(("pickup",c("b"))),var(("ontable",c("b"))),
    value(var(("ontable",c("b"))),true)).
pre(action(("pickup",c("b"))),var(("clear",c("b"))),
    value(var(("clear",c("b"))),true)).
post(action(("pickup",c("b"))),
    effect(unconditional),var(("holding",c("b"))),
    value(var(("holding",c("b"))),true)).
post(action(("pickup",c("b"))),
    effect(unconditional),var("handempty"),
    value(var("handempty"),false)).
post(action(("pickup",c("b"))),
    effect(unconditional),var(("clear",c("b"))),
    value(var(("clear",c("b"))),false)).
post(action(("pickup",c("b"))),
    effect(unconditional),var(("ontable",c("b"))),
    value(var(("ontable",c("b"))),false)).

action(action(("pickup",c("a")))).
pre(action(("pickup",c("a"))),var("handempty"),
    value(var("handempty"),true)).
pre(action(("pickup",c("a"))),var(("ontable",c("a"))),
    value(var(("ontable",c("a"))),true)).
pre(action(("pickup",c("a"))),var(("clear",c("a"))),
    value(var(("clear",c("a"))),true)).
post(action(("pickup",c("a"))),
    effect(unconditional),var(("holding",c("a"))),
    value(var(("holding",c("a"))),true)).
post(action(("pickup",c("a"))),
    effect(unconditional),var("handempty"),
    value(var("handempty"),false)).
post(action(("pickup",c("a"))),
    effect(unconditional),var(("clear",c("a"))),
    value(var(("clear",c("a"))),false)).
post(action(("pickup",c("a"))),
    effect(unconditional),var(("ontable",c("a"))),
    value(var(("ontable",c("a"))),false)).alse)).
post(action(("pickup",c("a"))),effect(unconditional),
var(("ontable",c("a"))),value(var(("ontable",c("a"))),false)).

goal(var(("ontable",c("a"))),value(var(("ontable",c("a"))),true)).
goal(var(("on",c("b"),c("a"))),value(var(("on",c("b"),c("a"))),true)).
\end{lstlisting} 

\end{document}